\documentstyle[twocolumn,epsfig]{elsart}
\begin{document}
\begin{frontmatter}
\title{Beam Test Results of the BTeV Silicon Pixel Detector}
\author[Fermi]{G. Chiodini},
\author[Fermi]{J.A. Appel},
\author[Syracuse]{M. Artuso},
\author[Fermi]{J.N. Butler}, 
\author[Fermi]{G. Cardoso}, 
\author[Fermi]{H. Cheung}, 
\author[Fermi]{D.C. Christian},
\author[INFN]{A. Colautti}, 
\author[Milan]{R. Coluccia},
\author[Milan]{M. Di Corato},
\author[Fermi]{E.E. Gottschalk}, 
\author[Fermi]{B.K. Hall}, 
\author[Fermi]{J. Hoff}, 
\author[Fermi]{P. A. Kasper},
\author[Fermi]{R. Kutschke}, 
\author[Fermi]{S.W. Kwan}, 
\author[Fermi]{A. Mekkaoui},
\author[INFN]{D. Menasce}, 
\author[Iowa]{C. Newsom}, 
\author[INFN]{S. Sala},
\author[Fermi]{R. Yarema},
\author[Syracuse]{J.C. Wang}, and
\author[Fermi]{S. Zimmermann}
\address[Fermi]{Fermi National Accelerator Laboratory, Batavia, IL 60510, USA}
\address[Iowa]{University of Iowa, Iowa City, IA 52242, USA}
\address[INFN]{Sezione INFN di Milano, via Celoria 16 - 20133 Milano, Italy}
\address[Milan]{Universit\'{a} di Milano, Dipartimento di Fisica, 
via Celoria 16 - 20133 Milano, Italy}
\address[Syracuse]{Syracuse University, Syracuse, NY 1344-1130, USA}
\begin{abstract}
The results of the BTeV silicon pixel detector beam test 
carried out at Fermilab in 1999-2000 are reported.
The pixel detector spatial resolution has been studied
as a function of track inclination, sensor bias, and readout threshold.
\end{abstract}
\end{frontmatter}

\section{Introduction}
The BTeV collaboration has intensively beam-tested several
single chip silicon pixel detector prototypes and front-end readout chips,
in order to establish the basic parameters of the pixel sensors and
readout chips which will be used as 
the building blocks of the BTeV vertex detector\cite{Newsom}.
To study the pixel detector spatial resolution,
a reference silicon telescope was used to 
project the incident beam track to the pixel sensor under test.
Of particular interest was a comparison of the resolution 
obtained, using 8 bit and 2 bit charge information, 
for a variety of incident beam
angles (from 0 to 30 degrees). Moreover, the spatial resolution 
was studied as a function of sensor bias and readout threshold.

\section{Experimental setup}

\begin{figure}
\begin{center}
\epsfig{figure=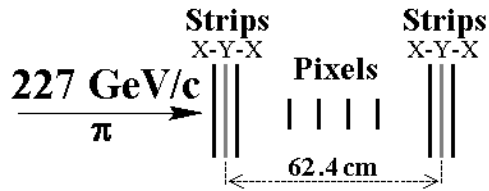,width=64mm,angle=0}
\end{center}
\caption{Schematic drawing of the silicon telescope.
\label{telescope} }
\end{figure}

The tests were performed in the MTest beamline at Fermilab, 
with a 227 GeV/c pion beam incident on a 6 plane silicon microstrip telescope
(see Figure~\ref{telescope}), with
several single-chip silicon pixel planes placed in the middle 
of the apparatus.
The pixel sensors tested have 50 $\mu$m $\times$ 400 $\mu$m pixel size and 
are all from the ``first ATLAS prototype submission
\cite{ATLAS_1st}''; up to four pixel detectors could be tested simultaneously.
The silicon microstrip detectors (SSD) were read out using
SVX-IIb ASIC's\cite{SVX-IIb} and
the data acquisition system was based on VME, 
adapted from the CDF SVX test stand\cite{CDF_teststand}.
The extrapolation accuracy of the silicon microstrip telescope at the pixel
detectors location was $\sim$ 2.1 $\mu$m for tracks with shared
charge in adjacent SSD channels. The excellent spatial resolution was due to
the small strip pitch (20 $\mu$m) and the high pion momentum available (which 
minimized the multiple scattering).

The readout was triggered by the coincidence of signals from two 15 cm $\times$
15 cm scintillation counters, positioned upstream and downstream of the silicon
telescope  and separated from each other by about 10 m. In order to select tracks
incident on the active area of the pixel detectors, the FAST\_OR
output signal from one of the FPIX0-instrumented pixel detectors
was also required.

\subsection{FPIX0 and FPIX1 readout chips}
\begin{figure}
\begin{center}
\epsfig{figure=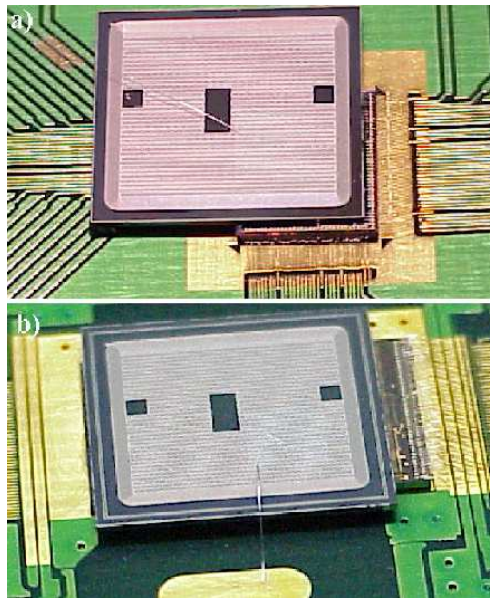,width=64mm,angle=0}
\end{center}
\caption{FPIX0 bonded to the CiS p-stop sensor (a) and 
FPIX1 bonded to the Seiko p-stop sensor (b).
\label{Fpix0and1} }
\end{figure}

The FPIX0 readout chips were indium bump bonded by Boeing North America Inc.
to CiS sensors (one p-stop ``ST1'' and one p-spray ``ST2'').
The instrumented portion of the sensor is 11 columns $\times$ 64 rows 
(Figure~\ref{Fpix0and1}a).
Each FPIX0 readout pixel contains an amplifier, a comparator, and a peak sensing
circuit\cite{Fpix}.
The analog output is 
digitized by an external 8-bit flash ADC.

The FPIX1 readout chips were indium bump bonded by Advanced Interconnect Technology Ltd
to Seiko sensors (two p-stop ST1's and one p-spray ST2).
FPIX1 is the first implementation of a high speed readout 
architecture designed for BTeV. 
It has 18 columns of 160 rows, and is the same size as the
ATLAS single chip sensors (Figure~\ref{Fpix0and1}b).  
However, a minor design error limited the number of rows
which may be read out to $\sim$ 90 per column. 
Each FPIX1 cell contains an amplifier, very similar to the FPIX0 amplifier, and 
four comparators, which form an internal 2-bit flash ADC\cite{Fpix}.

The pixel detectors were calibrated using a pulser and two x-ray sources 
(Tb and Ag foils excited by an $^{241}$Am $\alpha$-emitter).
For most of the data taking, the discriminator threshold for the FPIX0 p-stop
was set to a voltage
equivalent to 2500$\pm$400 e$^{-}$.
For the FPIX0 p-spray device the
corresponding threshold was typically 2200$\pm$350 e$^{-}$.
The amplifier noise was measured to be 105$\pm$15 e$^{-}$
for the FPIX0 p-stop
sensor.  The corresponding noise values for the FPIX0 p-spray sensor was 80
$\pm$10 e$^{-}$.
In addition, we found an equivalent charge noise due to the external
buffer amplifier and ADC of 400$\pm$150 e$^{-}$ for FPIX0 p-stop. 
The corresponding external noise values for the FPIX0 p-spray sensor were 185$\pm$20
e$^{-}$.
The FPIX1 chips have four threshold inputs
(one for each comparator in
the 2-bit FADC implemented in every cell).
We found a set of four average threshold values in nominal running conditions,
for the FPIX1 p-stop, 
of about $3780 e^{-}, 4490 e^{-}, 10290 e^{-},$ and $14680 e^{-}$, 
with a spread of about $380 e^{-}$.  The amplifier noise was measured to be
$110 \pm 30 e^{-}$. The relatively high FPIX1 readout threshold in the test beam 
was due to noise and pickup problems in a printed circuit board interface. 
An FPIX1 test module, with up to 5 chips bump bonded to an ATLAS tile-1 sensor,
has been operated stably in bench tests 
with discriminator threshold set below 1500 e$^-$.
 
\section{Results}

\subsection{Charge collection}

\begin{figure}
\begin{center}
\epsfig{figure=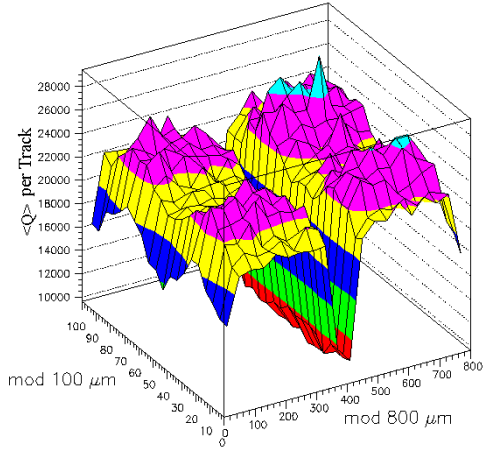,width=64mm,angle=0}
\end{center}
\caption{
The average pulse height versus track position
for the  CiS p-spray sensor bump-bonded to an FPIX0.
\label{losspspray2} }
\end{figure}

\begin{figure}
\begin{center}
\epsfig{figure=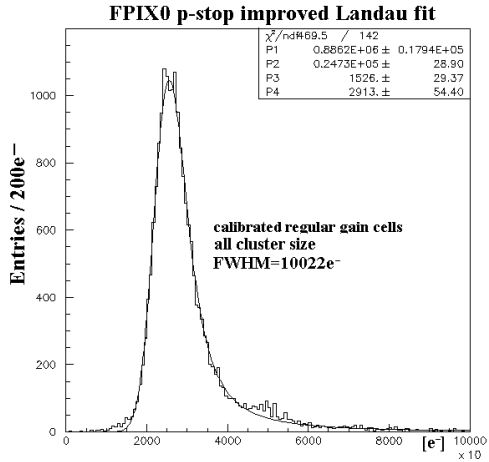,width=64mm,angle=0}
\end{center}
\caption{Pulse height distribution for  
the CiS p-stop sensor bump-bonded to an FPIX0.
\label{losspstop2}}
\end{figure}

Charge collection can be studied in detail for the FPIX0-instrumented sensors,
thanks to the 8-bit analog information and the absolute calibration.
Figure~\ref{losspspray2} shows that the p-spray sensor suffers sizeable
charge collection inefficiency between columns, especially on the
column boundaries
which include the ``punch-through biasing'' network\footnote{These charge 
losses are thought not to be intrinsic to the p-spray technology,
but a feature of this particular sensor design,
where each biased n$^+$ implant pixel is surrounded by a 
floating n$^+$ implant ring. The charge collection inefficiency
is believed to be mostly due to the presence of this ring\cite{Ragusa}.
}.
Our measurement of this charge loss is consistent
with previous measurements made by the ATLAS pixel collaboration\cite{Ragusa}. 
The measured pulse height distributions were fit using
a Landau function convoluted with a Gaussian~\cite{Hancock}.
Figure ~\ref{losspstop2}
shows the pulse height distributions for the FPIX0-instrumented p-stop sensor.
The ``improved Landau'' function fits the experimental data quite well, except
the bump at $\sim$ 50000 e$^-$ which is due to saturation of
the off-chip buffer amplifier/ADC combination. 
In addition, about 0.7\% of the events have a charge collected in the FPIX0 p-stop sensor 
less than 15000 e$^{-}$ (for
these values the Landau distribution  predicts a very small probability),
indicating that 
the p-stop sensor also suffers a very small
charge collection inefficiency. Studies show 
that this inefficiency is concentrated at
the four corners of the sensor pixels.

We find that the charge collected by the CiS p-spray sensor is about 24\% less than the
charge collected by the p-stop sensor.
The most probable and average charge collected are respectively 20000$\pm$70~e$^{-}$ 
and 23100$\pm$70~e$^{-}$ for the
CiS p-spray sensor (considering only tracks far away from the inter-pixel boundary).
For the CiS p-stop sensor, the most
probable charge collected is 24730$\pm$30~e$^{-}$, and the average charge
collected is 30100$\pm$30~e$^{-}$.

\subsection{Spatial Resolution}

\begin{figure}
\begin{center}
\epsfig{figure=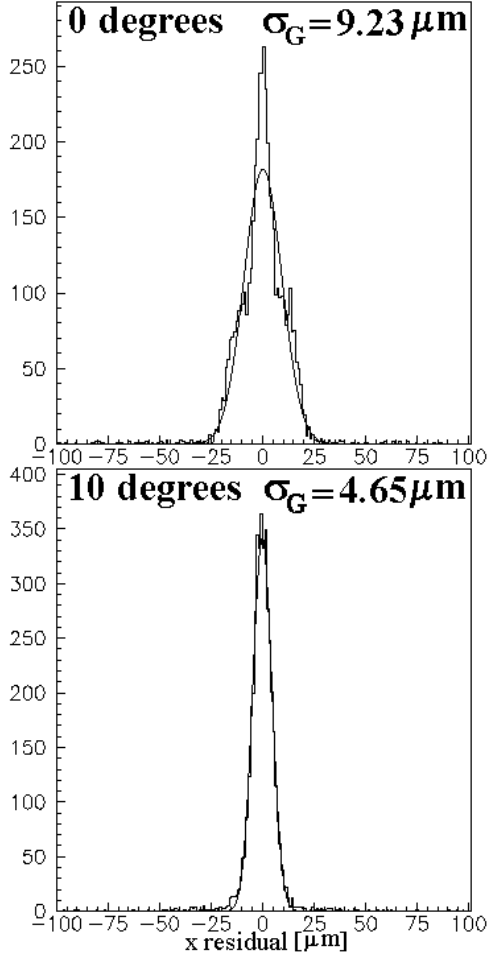,width=64mm,angle=0}
\end{center}
\caption{Residual distributions for the FPIX0 p-stop detector
for data taken with V$_{bias}=$-140V, Q$_{th}=$ 2500 e$^-$
and two different track inclinations.
\label{pstopstdres}}
\end{figure}

The tracks used to study pixel spatial resolutions were fit using data from the SSD
telescope and from pixel detectors other than the device under test, using
a Kalman-filter.
The coordinate measured by a pixel detector is obtained by the position of the
center of the cluster of hit pixels associated with a track, plus a correction
(conventionally called the $\eta$ function) which is a function of the charge
sharing, the cluster width, and the track angle.  For this analysis, we have used
a linear ``head-tail'' algorithm for computing the $\eta$ function, 
which ignores the charge deposited in pixels in the interior of a cluster, 
and uses only the charge deposited on the edges of the cluster\cite{Turchettapix}.
Two specific examples of residual distributions
are shown in Figure~\ref{pstopstdres} with the Gaussian fits superimposed.
Clearly, the residual distributions are not Gaussian, especially at zero degrees
where for a significant fraction of the time
only one pixel is hit.  
The residual distributions also have more entries far from zero
than the Gaussian fits. 
This can be clearly seen in the data taken
at ten degrees, when there is always charge sharing.  
The origin of these ``tails'' is attributed to the emission of $\delta$-rays which
skew the charge sharing and degrade the resolution.
Nonetheless, the Gaussian standard deviations provide a reasonably good
characterization of the width of the central peak for both plots.

\begin{figure}
\begin{center}
\epsfig{figure=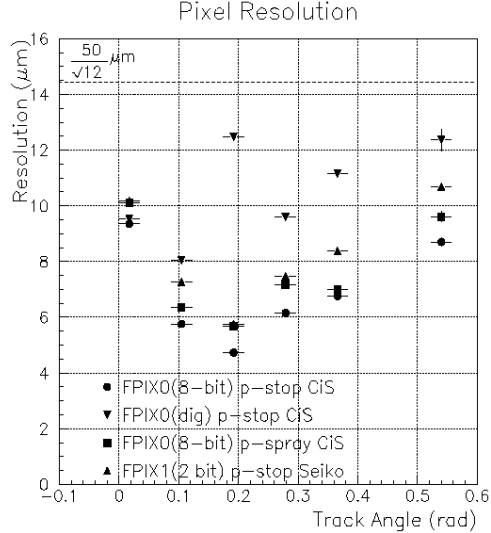,width=64mm,angle=0}
\end{center}
\caption{Position resolution along the short pixel dimension 
as a function of beam incidence angle for several detectors.
\label{pstopmod}}
\end{figure}

The residual distribution widths, obtained for several track angles and different
detectors, are shown in Figure~\ref{pstopmod}. The experimental 
results are in good agreement with the simulation results described in \cite{Artuso}.  
We have also computed
residual distributions for this data set without using any
charge sharing information.  
These ``digital'' resolution results are included in
Figure~\ref{pstopmod}.
The resolution for the FPIX1-instrumented p-stop detector
is slightly worse than the results that we obtained by degrading by software
the FPIX0-instrumented p-stop pulse height information to 2-bit equivalents.
This is because the main effect degrading the resolution is the high threshold
and the 2-bit analog information has only a minor effect.
In fact, the FPIX1-instrumented detector was operated
with a discriminator threshold of $\sim$3780 e$^-$, while the FPIX0-instrumented
detector was operated with a discriminator threshold of $\sim$2500 e$^-$.
Moreover, the results obtained for a p-spray detector with 
a threshold of $\sim$2200 e$^-$ show the extent to which the charge losses in
the p-spray sensor degrade the spatial resolution.

\begin{figure}
\begin{center}
\epsfig{figure=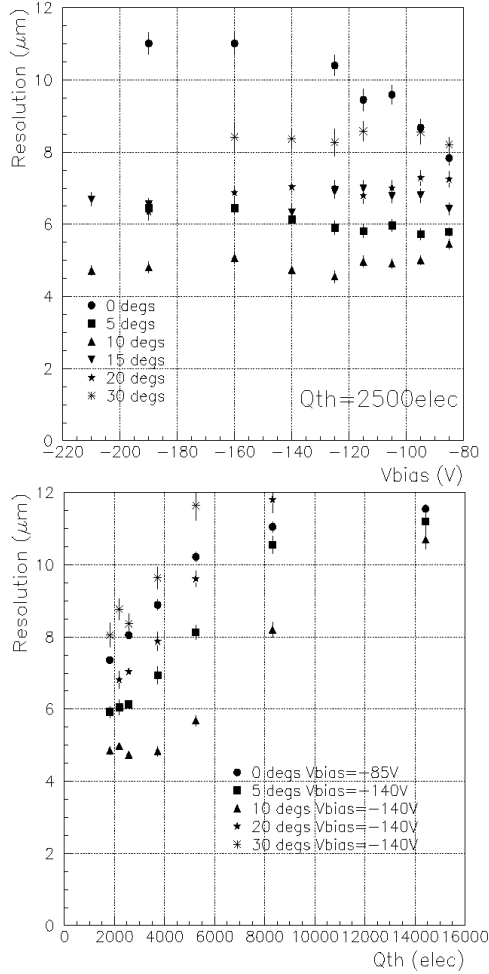,width=64mm,angle=0}
\end{center}
\caption{Spatial resolution versus bias voltage (upper plot) and
versus readout threshold (lower plot).
The data are from FPIX0-instrumented p-stop sensor with
a depletion voltage of 85V. 
\label{biasthrres}}
\end{figure}

Figure~\ref{biasthrres} shows how the position resolution is affected by changes
in the sensor bias voltage and the discriminator threshold.
The relative sensitivity of the pixel sensor position resolution
to these parameters is very important. In fact,
the variation of the bias
and effective threshold are similar to what is expected when 
the radiation damage influences the sensor bulk properties 
and the charge collection efficiency.
The data show, for large track angles, not too much sensitivity to the bias voltage,
because the charge-sharing is dominated by the track inclination.
At small track angle, when the diffusion gives a substantial contribution to the
charge-sharing, the sensor bias is important.
The effect of the readout threshold is 
always significant, 
but the spatial resolution is still better than 
$10 \mu m$ up to a threshold of $4000 e^{-}$.

\subsection{Resolution function shape}
\label{resshape}
\begin{figure}
\begin{center}
\epsfig{figure=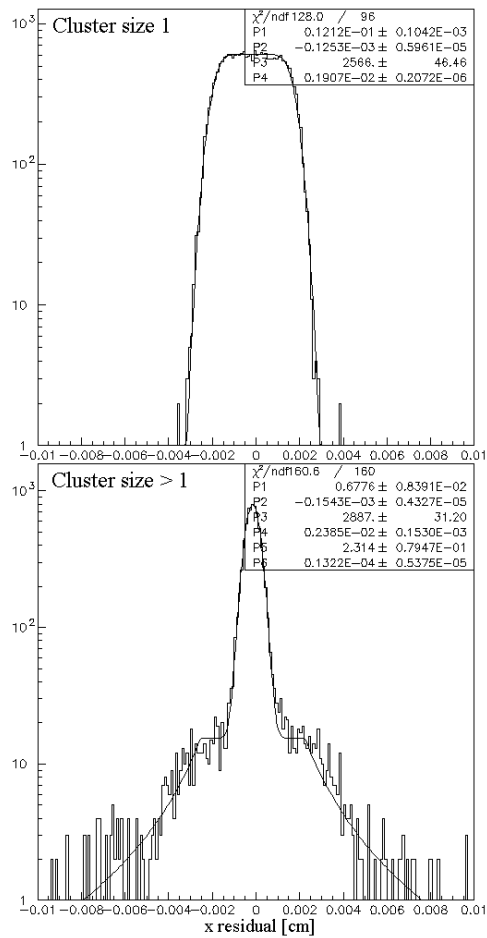, width=64mm, angle=0}
\end{center}
\caption{
Experimental residual distributions
fitted as described in sec.~\ref{resshape}. 
\label{tail}}
\end{figure}

The pixel residual distribution (or resolution function) 
deviates from a Gaussian in two important ways.
First, when tracks pass through one pixel only the 
residual distributions are well fitted
by a square function convoluted with a Gaussian.
Second, when tracks pass through more pixels
we have found that our experimental residual
distributions can be fitted 
by the sum of a Gaussian term $F_{G}$ 
and a term $F_{NG}$ which is a square
with edges that decrease like a power of 1/x:
\begin{equation}
 F_{NG}(x)= \left\{
 \begin{array}{c}
 \frac{A_{pl}}{|r_{c}|^{\gamma}} for |x|<r_{c}\\
 \frac{A_{pl}}{|x|^{\gamma}} for |x|>r_{c}
 \end{array}
 \right.,
\label{powerlaw}
\end{equation} 
\noindent
where $A_{pl}$ is a normalization constant, $r_{c}$ is the half width 
of the constant term, and $\gamma$ is the exponent of the power law.
Figure~\ref{tail} shows the experimental resolution functions
in log scale taken with the beam nominally at normal incidence for
cluster size one and cluster size bigger than one.  
In this last case, we find a satisfactory representation of the data 
using eq.~\ref{powerlaw} with the following set of parameters:
$\gamma = 2.3$, $r_{c} = 23.8 \mu$m, and $A_{pl}$ set so that $F_{NG}$ accounts
for about 18\% of the total number of entries in the distribution.

\section{Summary}
We have described the results of 
the BTeV silicon pixel detector beam test.
The pixel detectors under test used samples of the
first two generations of Fermilab pixel readout chips, FPIX0 and FPIX1,
(indium bump-bonded to ATLAS sensor prototypes).
The spatial resolution achieved using 
analog charge information is excellent
for a large range of track inclination. 
The resolution is still
very good using only 2-bit charge information.    
A relatively small dependence of the 
resolution on bias voltage is observed.
The resolution is observed to depend dramatically
on the discriminator threshold,
and it deteriorates rapidly for threshold 
above $4000e^{-}$.

\end{document}